%% file: main.tex
\documentclass[sigconf]{acmart}
\AtBeginDocument{%
  }

\copyrightyear{2026}
\acmYear{2026}
\setcopyright{cc}
\setcctype{by}
\acmConference[ICSE '26]{2026 IEEE/ACM 48th International Conference on Software Engineering}{April 12--18, 2026}{Rio de Janeiro, Brazil}
\acmBooktitle{2026 IEEE/ACM 48th International Conference on Software Engineering (ICSE '26), April 12--18, 2026, Rio de Janeiro, Brazil}
\acmPrice{}
\acmDOI{10.1145/3744916.3787764}
\acmISBN{979-8-4007-2025-3/2026/04}

\input{utils/macros}
\acmSubmissionID{icse2026-p1047}

\begin{document}
\title{ 
Enabling Global, Human-Centered Explanations for LLMs: \\From Tokens to Interpretable Code and Test Generation
}

\settopmatter{authorsperrow=3}
\author{Dipin Khati}
\authornote{Authors have contributed equally.}
\email{dkhati@wm.edu}
\affiliation{%
  \institution{William \& Mary}
  \city{Williamsburg}
  \state{Virginia}
  \country{USA}
}

\author{Daniel Rodriguez-Cardenas}
\authornotemark[1]
\email{dhrodriguezcar@wm.edu}
\affiliation{%
  \institution{William \& Mary}
  \city{Williamsburg}
  \state{Virginia}
  \country{USA}
}

\author{David N. Palacio}
\authornotemark[1]
\email{davidnad@microsoft.com}
\affiliation{%
  \institution{Microsoft}
  \city{Redmond}
  \state{Washington}
  \country{USA}
}

\author{Alejandro Velasco}
\email{svelascodimate@wm.edu}
\affiliation{%
  \institution{William \& Mary}
  \city{Williamsburg}
  \state{Virginia}
  \country{USA}
}

\author{Michele Tufano}
\email{tufanomichele@google.com}
\affiliation{%
  \institution{Google}
  \city{Kirkland}
  \state{Washington}
  \country{USA}
}

\author{Denys Poshyvanyk}
\email{dposhyvanyk@wm.edu}
\affiliation{%
  \institution{William \& Mary}
  \city{Williamsburg}
  \state{Virginia}
  \country{USA}
}

\renewcommand{\shortauthors}{Khati, Rodriguez-Cardenas, Palacio, et al.}

\begin{abstract}
As Large Language Models for Code (\lmc) become integral to software engineering, establishing trust in their output becomes critical. However, standard accuracy metrics obscure the underlying reasoning of generative models, offering little insight into how decisions are made. Although post-hoc interpretability methods attempt to fill this gap, they often restrict explanations to local, token-level insights, which fail to provide a developer-understandable global analysis. Our work highlights the urgent need for \textbf{global, code-based} explanations that reveal how models reason across code. To support this vision, we introduce \textit{code rationales} (\codeRational), a framework that enables global interpretability by mapping token-level rationales to high-level programming categories. Aggregating thousands of these token-level explanations allows us to perform statistical analyses that expose systemic reasoning behaviors. We validate this aggregation by showing it distills a clear signal from noisy token data, reducing explanation uncertainty (Shannon entropy) by over 50\%. Additionally, we find that a code generation model (\codeparrot) consistently favors shallow syntactic cues (e.g., \textbf{indentation}) over deeper semantic logic. Furthermore, in a user study with 37 participants, we find its reasoning is significantly misaligned with that of human developers. These findings, hidden from traditional metrics, demonstrate the importance of global interpretability techniques to foster trust in \lmc.
\end{abstract}

\ccsdesc[500]{Software and its engineering~Software development techniques~Automatic programming}
\ccsdesc[300]{Software and its engineering~Software development techniques~Program synthesis}
\ccsdesc[100]{Computing methodologies~Artificial intelligence}

\keywords{Interpretability, Trustworthy AI, LLMs for Code, Rationales, Explainability}



\maketitle

\input{text/0_intro_new}
\input{text/1_motivation}

\input{text/2_background}
\input{text/4_research_questions}
\input{text/5_methodology}
\input{text/6.User_study}
\input{text/6_rq1_result}
\input{text/7_rq2_result}

\input{text/9_related_work}
\input{text/10_lessons_learned}


\bibliographystyle{ACM-Reference-Format-num}
\bibliography{utils/rationales_bib}


\end{document}

%% file: utils/macros.tex
\usepackage{tikz}
\usepackage{caption}
\usepackage{blindtext}
\usepackage{tcolorbox}
\usepackage[final]{pdfpages}
\usepackage{lipsum,multicol}
\usepackage{xcolor}
\usepackage{tikz}
\usepackage{listings}
\usepackage{enumitem}
\usepackage{hyperref}
\usepackage{amsfonts}
\usepackage{wrapfig}
\usepackage{subcaption} 
\usepackage{adjustbox}
\usepackage{colortbl}
\usepackage{fancybox}
\usepackage{multirow}
\usepackage[normalem]{ulem}
\useunder{\uline}{\ul}{}
\usepackage{enumitem}
\usepackage{placeins}

\newcommand{\codeRational}{\textbf{Code}\ensuremath{\mathbb{Q}}\xspace}

\newcommand{\ie}{\textit{i.e.,}\xspace}
\newcommand{\eg}{\textit{e.g.,}\xspace}

\newcommand{\etal}{et al.\xspace}

\newtcolorbox{boxK}{
    fontupper = \small,
    sharpish corners, 
    boxrule = 0pt,
    toprule = 0pt, 
}

\newcommand{\secref}[1]{Sec.~\ref{#1}\xspace}
\newcommand{\figref}[1]{Fig.~\ref{#1}\xspace}
\newcommand{\tabref}[1]{Table~\ref{#1}\xspace}

\newcommand*\circled[1]{\tikz[baseline=(char.base)]{
            \node[shape=circle,draw,inner sep=0.5pt] (char) {#1};}}

\newcommand{\codeparrot}{\textit{codeparrot-small}\xspace}
\newcommand{\bart}{\textit{BART-large}\xspace}

\newcommand{\sgbd}{$TB_1$\xspace}
\newcommand{\dcsgbd}{$TB_2$\xspace}
\newcommand{\dcsg}{$TB_3$\xspace}
\newcommand{\dc}{$TB_4$\xspace}

\newcommand{\levelone}{\textbf{$L_1$}\xspace}
\newcommand{\leveltwo}{\textbf{$L_2$}\xspace}
\newcommand{\levelthree}{\textbf{$L_3$}\xspace}

\newcommand{\lmc}{LM4Code\xspace}
\newcommand{\lmsc}{LM4Code\xspace}

\newcommand{\ewash}{\textit{eWASH}\xspace}

\newcommand{\aise}{AI4SE\xspace}

\newcommand{\statements}{\texttt{\small{[statements]}}\xspace} 
\newcommand{\structural}{\texttt{\small{[structural]}}\xspace} 
 
\newcommand{\oop}{\texttt{\small{[oop]}}\xspace} 
\newcommand{\conditionals}{\texttt{\small{[conditionals]}}\xspace} 
\newcommand{\loops}{\texttt{\small{[loops]}}\xspace} 
\newcommand{\asserts}{\texttt{\small{[asserts]}}\xspace} 
\newcommand{\punctuation}{\texttt{\small{[punctuation]}}\xspace} 
\newcommand{\class}{\texttt{\small{[class]}}\xspace} 
\newcommand{\focalmethod}{\texttt{\small{[focal method]}}\xspace} 
\newcommand{\signature}{\texttt{\small{[signature]}}\xspace} 
\newcommand{\bool}{\texttt{\small{[bool]}}\xspace} 
\newcommand{\expression}{\texttt{\small{[expression]}}\xspace} 
\newcommand{\nlnoun}{\texttt{\small{[nl\_noun]}}\xspace} 
\newcommand{\exceptions}{\texttt{\small{[exceptions]}}\xspace} 
\newcommand{\functional}{\texttt{\small{[functional]}}\xspace} 
\newcommand{\return}{\texttt{\small{[return]}}\xspace} 

%% file: text/0_intro_new.tex
\section{Introduction}\label{sec:introduction}

\lmc such as GPT-4\cite{brown2020languagemodelsfewshotlearners}, Claude 3\cite{TheC3}, and StarCoder\cite{li2023starcodersourceyou} promise significant developer productivity as developers integrate them into their software engineering (SE) workflows. As these models move from research curiosities to indispensable tools for tasks such as code completion and test generation, their reliability, safety, and \textit{interpretability} become paramount. The central question for the AI for Software Engineering (\aise) community is no longer ``Can these models generate \textit{correct} code?'' but rather ``Can we \textit{trust} generative models?"~\cite{khati2025mappingtrustterrainllms}. Answering this question demands that we scrutinize the underlying reasoning of generative models (\ie LLMs), which aligns with calls for a more rigorous science of interpretable machine learning~\cite{doshi-velez_towards_2017, palacio_toward_2024}.

The primary obstacle to establishing code-based trust lies in our current evaluation paradigm. The \textit{de facto} standard for evaluating LLMs consists of a suite of accuracy-based metrics, including \textit{Exact Match} and \textit{F1} for general evaluation; CodeBLEU for code synthesis~\cite{ren2020codebleumethodautomaticevaluation}; HumanEval, with \textit{pass@k} and correctness rate, for code reasoning~\cite{chen_evaluating_2021}. Although these metrics possess useful information for measuring functional correctness, they offer no insight into the model's decision-making process. For instance, a model might generate correct code by relying on spurious, non-generalizable patterns in the input, a behavior that accuracy scores alone cannot detect~\cite{Ribeiro2020BeyondAB}. Overreliance on correctness can overestimate the true capabilities of the model~\cite{burnell_rethink_2023} and fails to meet the growing demand for a more holistic evaluation of LLMs~\cite{liang2023holisticevaluationlanguagemodels}.

To look beyond precision, the AI community has increasingly turned to post-hoc interpretability techniques such as LIME~\cite{ribeiro_why_2016} and Shapley values~\cite{lundberg_unified_nodate} that can explain individual predictions by attributing importance to input features. Although valuable, these methods are often computationally intensive, conceptually ill-suited for generative tasks, and provide only local, token-level explanations. Similarly, attention mechanisms initially offered a promising window into model reasoning~\cite{mohankumar_towards_2020}, however, subsequent work challenged their reliability, showing that attention does not accurately represent the internal logic of a model~\cite{jain_attention_2019, serrano_is_2019}. Consequently, practitioners lack tools to systematically analyze a model's behavior at a global level to uncover consistent biases or diagnose systemic reasoning flaws and successes.

To address these limitations, we introduce \codeRational, a \textit{interpretability framework} that enables global post-hoc analysis of LLMs for code. Our framework was inspired by rationalization theory, identifying the minimal subset of input tokens that are the most influential for a prediction~\cite{vafa_rationales_2021}. However, for explanations to be truly useful to practitioners, they must be presented in human-understandable terms~\cite{doshi-velez_towards_2017}. Explanations at the raw token level are often too fine-grained and lack the abstract structure that developers use to reason about programs~\cite{Ghorbani19}. Our key contribution, therefore, is a framework that goes beyond tokens by mapping these low-level rationales to a category of higher-level programming categories derived from syntax decomposition~\cite{palacio2024trustworthyinterpretablellmscode}. This abstraction of individual tokens into meaningful categories (\eg from \texttt{`if'} to \conditionals) is the critical step that allows us to aggregate explanations across thousands of outputs. Our research enables a \textit{global} statistical view (\ie reasoning patterns) of a given generative model (\eg encoder-decoder and decoder-only), directly addressing the limitations of previous local methods (\eg Shapley, SHAP, and Lime).

To establish and rigorously validate our framework, we focus on two well-known models representing distinct decoder-only (\codeparrot) and encoder-decoder (\bart) architectures for Python code completion and Java test case generation, respectively. This work is outside the scope of benchmarking specific state-of-the-art models; instead, we demonstrate the flexibility and power of our transformer-agnostic \codeRational framework and highlight the need for practitioner-friendly global explanations.

The global statistical view of our study yields immediate and telling insights into the LLM's behavior. Applying our framework, we find that LLMs systematically rely on shallow syntactic patterns, such as code \textbf{indentation} and \textbf{punctuation}, far more frequently than on deep semantic categories, such as conditional \textbf{logic}. We also observe a strong \textbf{self-reliance} phenomenon in which structures like \texttt{loops} are primarily influenced by existing loop-related tokens in the prompt. Perhaps most critically, our user study of $37$ participants reveals a significant misalignment between these machine-generated rationales and the explanations provided by human developers (Jaccard similarity 0.074 for code completion, 0.306 for test generation), highlighting a fundamental gap in their respective reasoning processes. These findings, invisible to accuracy metrics, underscore the need for category-level global interpretability to foster appropriate trust and guide future model development.

The main contributions of our work are as follows: (1) We developed an interpretability framework \codeRational, which transforms token rationales at the low level into a structured interpretability tensor based on high-level programming concepts; (2) We performed an \textit{exploratory analysis} using two different models to demonstrate the \textit{applicability} of \codeRational to estimate global code-based explanations; (3) We conducted a \textit{user study} to assess \textit{usability} of \codeRational, where participants found it informative, readable, and useful to show statistical correlations and interpret code predictions; our study also revealed a gap between human reasoning and model reasoning, indicating directions for future work; (4) We published an online appendix~\cite{anonymized_repo} that contains documented notebooks for researchers, experimental data, source code, models, and the statistical analysis of the results of the user study.

%% file: text/1_motivation.tex
\section{Why code-based global explanations?}

\begin{figure}[t]
  \centering
  \vspace{-0.2cm}
  
  \includegraphics[width=\linewidth]{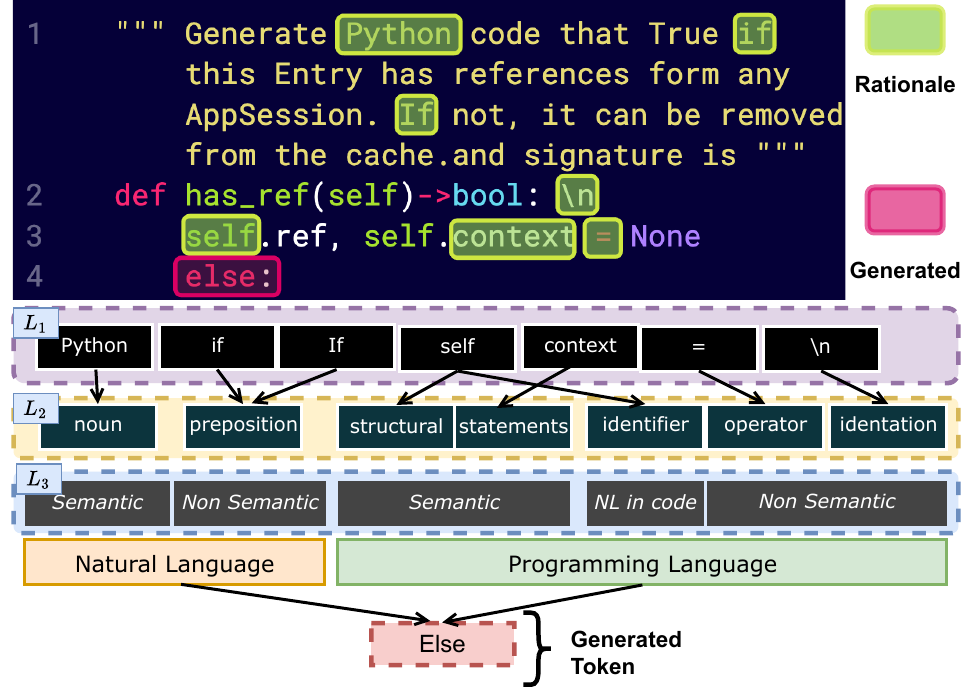}
  
  \vspace{-0.3cm}
  
  \caption{Conceptual Dependency Map}
  \label{fig:dependency_map}
  
  \Description{The figure shows a conceptual dependency map illustrating the CodeQ framework. It visualizes the hierarchy from raw tokens (L1) like 'Python', 'if', and 'self' to categories (L2) such as 'noun', 'preposition', and 'structural statements'. These are further grouped into semantic levels (L3). The diagram shows how Natural Language and Programming Language modalities converge to influence the generated token 'Else'.}
  
  \vspace{-0.5cm}
\end{figure}

Although researchers acknowledge the need for interpretability in \lmc, existing techniques provide limited utility to software developers. The AI4SE community focuses on \textit{ local}, instance-level explanations, which analyze a single prediction at a time. For example, as shown in ~\figref{fig:dependency_map}, a local method can explain why the model predicted \textit{``else''} (in red) by highlighting influential tokens in the prompt (in green). Despite this necessary first step, this isolated insight falls short of providing a global and comprehensive understanding required to trust or debug a model.

This focus on local analysis stems from the limitations of dominant \textit{post-hoc} approaches. Perturbation-based methods such as LIME\cite{ribeiro_why_2016} and SHAP\cite{lundberg_unified_nodate}, when applied to code classification tasks, require significant computation and do not scale to generative settings~\cite{limeshaplimit}. Attention-based explanations, although readily available in transformer models, have been shown by a significant body of work to unfaithfully represent the model's internal logic~\cite{jain_attention_2019, serrano_is_2019}. Even code-specific techniques, like counterfactual explanations, concentrate on local, instance-level changes. This methodological gap leaves practitioners without tools to form and rigorously test broad hypotheses about model behavior. Without a global view, a developer cannot falsify a claim such as ``the model consistently understands conditional logic,'' a key practice for building justified confidence in any system~\cite{lipton2017mythos}.

Even if these methods could scale globally, they would produce an incomprehensible result. Explanations at the raw token level, while faithful to the machine, overwhelm developers with fine-grained detail, leading to cognitive overload that makes extracting meaningful insights nearly impossible~\cite{poursabzi-sangdeh_manipulating_2021}. This problem arises because developers tend to reason about programs using abstract categories such as prepositions and operators, rather than individual tokens~\cite{fagerholm2022cognitionsoftwareengineeringtaxonomy}. The concept levels \levelone, \leveltwo, and \levelthree in ~\figref{fig:dependency_map} demonstrate this principle, mapping token rationales to developer-centered categories like ``statements'' provides a more insightful and actionable explanation than simply pointing to the token \textit{"self"}.

{To truly advance the trustworthiness of LLMs in SE, our interpretability frameworks must evolve. We must move beyond both the local scope of current methods and the machine-level granularity of token-based output. The core novelty of this paper is a framework \codeRational, which enables this conceptual leap. Although we build upon a token-level rationale method~\cite{vafa_rationales_2021}, our contribution is the \textit{abstraction and aggregation} of its noisy, local outputs (as we show in Sect.~\ref{result:distribution}) into the global, developer-centered explanations that form the language of software development itself.}

%% file: text/2_background.tex
\section{Background}
\label{sec:interpretability_background}

\textbf{Theory of (Code) Rationales.} We build on the sequential rationales framework by Vafa et al.~\cite{vafa_rationales_2021}, which identifies the minimal subset of input tokens (rationales) necessary to preserve the prediction of a model. This approach allows practitioners to extract interpretable token-level explanations for individual predictions in transformer models. Since computing the optimal rationale set is intractable, Vafa et al. adopt a greedy rationalization strategy, starting from an empty context and incrementally adding tokens that most increase prediction confidence.

To ensure meaningful behavior in incomplete inputs, the model must be \textit{compatible}, \ie capable of producing stable outputs when conditioned on partial contexts. This is achieved by fine-tuning the model on randomly sampled input subsets, as described in~\cite{vafa_rationales_2021}. In our work, we adopt this formulation and apply it to \lmc without modifying the underlying greedy algorithm or compatibility setup.

\textbf{Interpretability Desiderata.} To guide our analysis, we adopt three common interpretability desiderata: \textit{Trust}, \textit{Informativeness}, and \textit{Usability}, widely discussed in the literature~\cite{lipton2017mythos, doshi-velez_towards_2017, DesiderataSokol,khati2025mappingtrustterrainllms}. \textit{Trust} refers to the practitioner's confidence in the model's behavior, especially under uncertainty. \textit{Informativeness} emphasizes the need for explanations to provide useful and context-rich insights beyond simple token attribution. \textit{Usability} focuses on the effectiveness of the interpretability method in real-world workflows, ensuring that explanations are actionable and understandable by human developers. We argue that these desiderata are important for any interpretability framework.

\textbf{Context Window for Test Case Generation.} A \textit{focal method} is the specific method in a codebase that a test case is primarily intended to verify or exercise—it represents the ``method under test.'' A \textit{context window}, on the other hand, refers to the broader code environment surrounding the focal method (\textit{fm}), which may include the class name \textit{(fc)}, constructor signatures \textit{(c)}, other method signatures \textit{(ms)}, and public fields of the class \textit{(ff)}\cite{Tufano_2022}. This contextual information is crucial because it provides a richer and more complete view of how the focal method operates within its environment, which helps machine learning models generate more accurate and semantically meaningful test cases.

\begin{figure}[ht]
  \centering
  \vspace{-0.2cm}
  
  \includegraphics[width=0.9\linewidth]{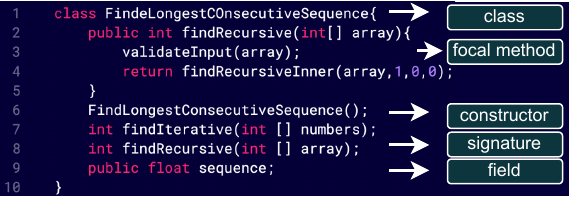}
  
  \vspace{-0.3cm}
  
  \caption{\ewash Focal Context Window for Test Generation}
  \label{fig:ewash_taxonomy}
  
  \Description{The figure illustrates the eWASH taxonomy applied to a Java class 'FindLongestConsecutiveSequence'. It annotates specific code regions with labels: 'class' for the class definition, 'focal method' for the method under test (findRecursive), 'constructor' for the class initialization, 'signature' for other method declarations, and 'field' for class variables.}
  
  \vspace{-0.5cm}
\end{figure}





%% file: text/4_research_questions.tex
\section{Research Questions}
We conducted an exploratory analysis and a user study to explore the following RQs: 

\begin{enumerate}[label=\textbf{RQ$_{\arabic*}$}, ref=\textbf{RQ$_{\arabic*}$}, wide, labelindent=5pt]\setlength{\itemsep}{0.2em}
    
    \item \label{rq:applicability}\textbf{[Applicability]:} \textit{How applicable is \codeRational to interpret code generation?} {We define applicability as the ability to use \codeRational in crafting understandable explanations.  This RQ focuses on exploring the experience of using \codeRational to explain the behavior of \lmsc in code generation tasks.  We hypothesize that through greedy rationalization, we can explain the most important rationales of an input that lead to a certain prediction for both local and global scenarios.}
    \begin{enumerate}[label=\textbf{RQ$_{1.\arabic*}$:}, ref=\textbf{RQ$_{1.\arabic*}$}, wide, leftmargin=0.2cm]\setlength{\itemsep}{0.2em}
       
        \item \label{rq:applicability_one}{\textit{What do the distributions of token-level rationale probabilities reveal about the interpretability of model decisions, and how does this inform the need for concept-based abstraction?}}
         \item \label{rq:applicability_two}{\textit{What are the most frequent rationales for code completion?}} 

         \item \label{rq:applicability_global}{\textit{How can concept-level rationale be used to provide global post hoc interpretability of code generation model?}} 

        \item \label{rq:applicability_three} {\textit{How does \codeRational perform when applied to different architectures in test-case generation scenario?}} 
        \end{enumerate}

     \item \label{rq:utility} \textbf{[Usability]:} \textit{How useful is  \codeRational in practical settings?} We validate the extent to which \codeRational is useful in practical settings in a user study. We measure the usability of our framework with qualitative metrics, such as the usefulness and readability of \codeRational, as well as the degree to which \codeRational helps assess the alignment of \lmsc.

\end{enumerate}

%% file: text/5_methodology.tex
\section{\codeRational Interpretability Framework}\label{sec:experiments}

This section outlines the \codeRational \textit{framework}, introducing preconditions and steps to generate local and global explanations. This section also includes the case study setup to evaluate the applicability of \codeRational \ref{rq:applicability}. 

\subsection{Interpretability Setup}
\label{sec:int_setup}
The \textit{\codeRational framework} enables practitioners to examine which parts of the prompt contribute the most to the generation of code prediction. The relationship between the prompt and the output can be evaluated for a snippet scope (\ie using dependency maps \figref{fig:dependency_map}) as a \textit{local explanation} or for a complete dataset in a \textit{global explanation}. \codeRational framework consists of four steps to generate a local and global explanation.

\begin{figure}[t]
  \centering
  \vspace{-0.2cm}
  
  \includegraphics[width=\linewidth]{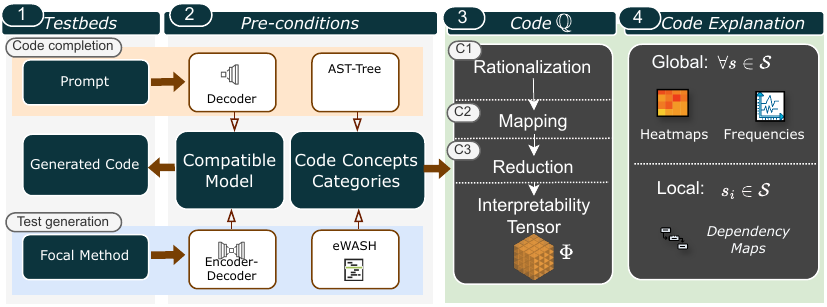}
  
  \vspace{-0.2cm}
  
  \caption{\codeRational Interpretability \textit{Framework}}
  \label{fig:framework}
  
  \Description{The figure illustrates the CodeQ framework workflow across four steps: (1) Testbeds preparation for code completion and test generation; (2) Pre-conditions establishment using compatible models and code concept categories; (3) The CodeQ core process involving Rationalization, Mapping, and Reduction to produce an Interpretability Tensor; and (4) Code Explanation generation, outputting global heatmaps and local dependency maps.}
  
  \vspace{-0.5cm}
\end{figure}

\begin{enumerate}[
    label={\textbf{Step$_{\arabic*}$:}}, 
    ref=Step$_{\arabic*}$, 
    wide, 
     labelindent=5pt
]
\setlength{\itemsep}{0.2em}

\item \label{step1} \textbf{Constructing Testbeds.} {To support meaningful analysis, the interpretability testbed must include diverse prompt types that preserve the semantic structure of the input-output interactions. Our curated testbed enables both local and global explanations.} 

For the code generation case, we mined 50K unique Python snippets using Galeras~\cite{Daniel23}. From this, we construct four prompt-based testbeds, each capturing a different context configuration: (1) signature + truncated body (\sgbd), (2) docstring + signature + body (\dcsgbd), (3) docstring + signature (\dcsg), and (4) docstring only (\dc). Each testbed contains 100 unique prompts. To support robust statistical analysis, we sampled 30 rationale sets for each prompt, obtaining a total of 12K model-generated sequences for code generation.
 
{For test case generation case, we used the \texttt{\small Methods2Test}~\cite{Tufano_2022} dataset. Each instance consists of a focal method and its corresponding JUnit test. We used $1K$ samples in total. This testbed enables us to investigate how encoder-decoder models handle test generation and identify which parts of the input Java context (\eg method body, fields, etc.) are most influential in generating relevant tests.}
\item \label{step2} \textbf{Preparing Model and Task Concepts.}  
To extract meaningful rationales from \lmsc, we begin by preparing two key components: a compatible model and a set of interpretable concepts $\mathcal{C}$ aligned with the software engineering task.

\textbf{Concept Extraction.}  
\lmsc operates at the token level; however, token-level rationales may not directly correspond to semantically meaningful units. To support human-centered interpretability, we must define a task-specific interpretable \textit{concepts} $\mathcal{C}$ that groups tokens into higher-level constructs such as control flow structures, method identifiers, and documentation tags. This enables aggregation and explanation of model behavior at a concept level.
\begin{figure}[t]
  \centering
  \vspace{-0.2cm}
  
  \includegraphics[width=\linewidth]{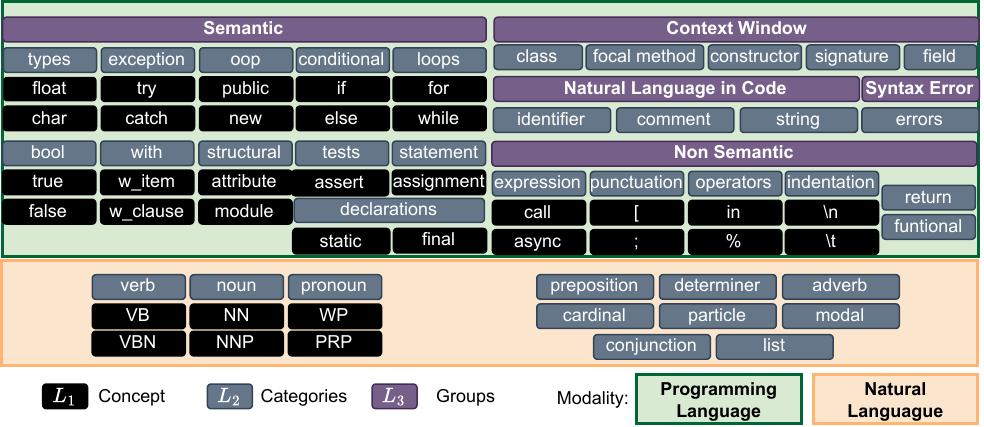}
  
  \vspace{-0.3cm}
  
  \caption{Interpretability Concepts ($\mathcal{C}$) for Java and Python }
    \vspace{-0.3cm}

  \label{fig:rationales_taxonomy}
  
  \Description{The figure presents the hierarchical taxonomy of interpretability concepts. It maps low-level concepts (L1) into higher-level categories (L2) and groups (L3). The taxonomy is divided into two modalities: Programming Language (containing Semantic groups like types, loops, conditionals, and Non-Semantic groups like punctuation) and Natural Language (containing Semantic groups like nouns, verbs, and Non-Semantic groups like prepositions).}
  
  \vspace{-0.1cm}
\end{figure}

{We propose two types of interpretability concepts tailored to different tasks: (1) For code completion, we extract Abstract Syntax Tree (AST)-based concepts using Tree-Sitter\cite{tree-sitter} and enrich them with natural language categories using NLTK (\eg identifiers, descriptions)\cite{nltk}, as shown in \figref{fig:rationales_taxonomy}. ASTs offer a natural, language-agnostic abstraction that captures the structural and semantic hierarchy of source code, aligning closely with how developers reason about programs; (2) For test generation, we adopt the focal method scoping strategy defined in \ewash~\cite{clement_long-range_2021}, which captures relevant Java class-level context—such as fields, constructors, and surrounding methods as shown in \figref{fig:ewash_taxonomy}.}

\textbf{Model Setup.}  
Sequential rationalization requires the model to make consistent predictions even when provided only partial inputs. To address this, we follow Vafa et al.~\cite{vafa_rationales_2021} and make the model \textit{compatible} by fine-tuning it on randomly masked input prefixes using word dropout.

{For code completion, we use \codeparrot~\cite{codeparrot}, a decoder-only Transformer with 110M parameters, pretrained on the \textit{codeparrot-clean} dataset (5M Python samples). To ensure compatibility, we fine-tune the model on the same dataset using a word dropout rate of 0.5, a learning rate of $1e\text{-}5$, and the Adam optimizer over 60K steps. We maintain all original architecture and hyperparameters.} {For test-case generation in Java, we use the \bart model implemented via Fairseq \cite{ott2019fairseqfastextensibletoolkit}, fine-tuned on the \texttt{Methods2Test} dataset \cite{Tufano_2022}. This encoder-decoder model (400M parameters) was trained with inverse square root learning rate scheduling, FP16 mixed precision, inverse-length word dropout, and reset optimizers.}

\item \label{step3} \textbf{Building Interpretability Tensors.}  
To support both local and global explanations, we convert token-level rationales into structured, concept-aligned tensors. This step involves three key components: (1) extracting rationales using greedy approximation, (2) mapping token-level rationales to human-understandable concepts, and (3) aggregating these into interpretable tensors that generalize across multiple sequences.

\textbf{Component$_1$: Rationalization.}  
We selected the greedy rationalization approach~\cite{vafa_rationales_2021} over perturbation-based baselines such as LIME~\cite{ribeiro_why_2016} and SHAP~\cite{lundberg_unified_nodate} for three primary reasons. First, LIME and SHAP are \textit{local} explainers suited to static predictions, making them incompatible with our goal of deriving \textit{global} insights from a \textit{dynamic, autoregressive} model. Their position-specific attributions do not aggregate reliably across time or across sequences. Second, these methods are computationally infeasible at our scale: KernelSHAP requires approximately $2p+2048$ model evaluations (e.g., $\approx 3{,}048$ calls for $p=500$ tokens) for one explanation, while the $O(t \cdot p)$ complexity of greedy rationalization, where $t$ is the number of tokens selected in the rationale, enables analysis of more than 12K sequences. Third, greedy rationalization provides \textit{faithful} explanations by design by selecting minimal sufficient subsets of tokens that preserve the model’s predictions, allowing principled aggregation into higher-level concepts. Perturbation-based methods do not guarantee such faithfulness, and attention-based explanations have been repeatedly shown to be unreliable for attributing model reasoning~\cite{jain_attention_2019, serrano_is_2019, ribeiro_why_2016}. 


To apply greedy rationalization~\cite{vafa_rationales_2021}, we extracted the smallest subset of input tokens (the rationale) that preserves the model’s prediction. Let \( w_{1:T} \) denote the input sequence of length \( T \), and let \( w_t \) be the token being predicted. The algorithm finds the rationale \( r \subseteq w_{<t} \) by building it token by token.

The rationale starts with an empty set (\( r^{(0)} = \emptyset \)) and, at each step \(k\), iteratively adds the single most influential token from the remaining context. This process is formalized in Equation~\ref{eq:greedy}. The subscript, \textbf{\( j \in \{[t-1] \setminus r^{(k)}\} \)}, defines the search space: it looks at every token \(j\) that is in the input context before the target (\([t-1]\)) but has \textbf{not} already been added to the rationale set (\(\setminus r^{(k)}\)). For each candidate token, the algorithm calculates the new probability of predicting \(w_t\). The token resulting in the highest probability is added to the rationale for the next step. In our example \figref{fig:dependency_map}, to explain \texttt{``else''} the rationalization step identified rationales \texttt{``Python'', ``if'',``If'', ``\textbackslash n'', ``self'', ``context'', and ``=''}.

\begin{equation}
r^{(k+1)} = r^{(k)} \cup \arg \max_{j \in \{[t-1] \setminus r^{(k)}\}} P_{\theta}(w_t \mid w_{r^{(k)} \cup \{j\}})
\label{eq:greedy}
\end{equation}

The output of this process for a single sequence is a binary rationale matrix \( \phi \in \mathbb{R}^{T \times T} \), where \( \phi_{ij} = 1 \) if token \( w_i \) was used to predict token \( w_j \). This matrix represents all direct token-to-token influences. To account for variability in autoregressive decoding, we repeat this process 30 times for each sequence in our testbeds, yielding 30 rationale matrices ($\phi^{(n)}$) per sequence.

\begin{figure}[ht]
 
\centering
\includegraphics[width=\linewidth]{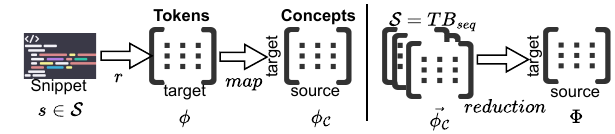}
\label{fig:interpretability-pipeline}
\vspace{-2em}
\caption{\codeRational Interpretability Tensors}
  \vspace{-0.3cm}

\end{figure}

\textbf{Component$_2$: Mapping.}  
To group token rationales into interpretable structures, we map each token in the rationale matrix \( \phi \) to a concept \( c \in \mathcal{C} \) (\levelone). This results in a matrix over the concept space:

\begin{equation}
\phi_\mathcal{C} = \phi_{[|\mathcal{C}| \times |\mathcal{C}|]} = \text{map}(\phi_{[T \times T]}, \mathcal{C})
\label{eq:mapping}
\end{equation}

Here, \( \phi_\mathcal{C} \in \mathbb{R}^{|\mathcal{C}| \times |\mathcal{C}|} \) is the \textit{concept-level rationale matrix}, where rows represent source concepts and columns represent target concepts. Each cell \( \phi_\mathcal{C}[i, j] \) reflects the presence or intensity of a rationale from concept \( c_i \) to concept \( c_j \).

{The resulting matrix \( \phi_\mathcal{C} \) is referred to as the \textit{interpretability matrix}. Since the mapping reduces the dimensionality from the original token space \( [T, T] \) to the concept space \( [|\mathcal{C}|, |\mathcal{C}|] \), it may not preserve all the detailed interactions present in the original rationale matrix \( \phi \). In particular, \( \phi_\mathcal{C} \) eliminates positional information, such as the exact sequence of tokens, and aggregates token-level signals into broader conceptual categories.}

This mapping component applied again over the  \( \phi_\mathcal{C} \), computes a broader level of concepts. For instance, \figref{fig:dependency_map} depicts some keywords (\eg \texttt{``self'', ``context''})  aggregated into categories (\leveltwo) such as \structural, \statements.  \codeRational allows practitioners to introduce concepts (\levelone), categories (\leveltwo), and groups (\levelthree) tailored to the use case, as shown in the \figref{fig:rationales_taxonomy} for code completion.  For test generation, we also introduced the focal concept categories (\figref{fig:ewash_taxonomy}) following the \ewash taxonomy for test case generation.



\textbf{Component$_3$: Reduction.}  
To support global analysis, we aggregate the rationale matrices across samples and trials. Let \( \Vec{\phi_\mathcal{C}} = \{\phi_\mathcal{C}^{(1)}, \phi_\mathcal{C}^{(2)}, \ldots, \phi_\mathcal{C}^{(N)}\} \) be a collection of concept-level rationale matrices from \( N \) trials. We define the reduction function as:

\begin{equation}
\Phi = reduce( \Vec{\phi_{C}} )=  \forall_{c} \in C, \forall_{s} \in S : g(\phi_{c}^{s})
 \label{eq:reduce}
\end{equation}

{This component focuses on summarizing interpretability information across multiple examples to support global analysis. The component applies the mapping in the previous \textit{Component$_2$}, in which we obtain an interpretability matrix \( \phi_\mathcal{C} \) for each snippet \( s \in \mathcal{S} \). These matrices reflect how different interpretability concepts relate within individual sequences. To analyze broader trends across the entire testbed, we combine them into a single global interpretability tensor \( \Phi \) using the reduction function \( g \).}

The function $g$ defines how values are aggregated, for example, by computing the average, maximum, or another summary statistic across sequences $S$ from the $N$ trials. The resulting tensor $\Phi$ has shape $[|\mathcal{C}|, |\mathcal{C}|]$ and captures the strength of conceptual relationships at a global level.
 
\item \label{step4} \textbf{Code-based Explanation.} 
The final step utilizes the interpretability tensors developed in previous steps to generate visual model explanations. These explanations can be local (\ie focused on individual model predictions) or global (\ie capturing high-level trends across a dataset).

Local explanations are derived from the interpretability matrix $\phi_\mathcal{C}$, which shows how concepts interact within a single input-output pair. These matrices can be visualized as dependency maps, which illustrate the influence of different tokens or concepts at various levels of abstraction. For example, \figref{fig:rationales_taxonomy} demonstrates three levels of interpretability according to the \textit{Components$_3$}: \levelone) concept-level rationales, \leveltwo) categories-level rationales, \levelthree) groups of categories (\eg context window). Explanations also include modalities, \ie modality-level attribution. 

Global explanations are built using the interpretability tensor $\Phi$ introduced in the reduction step. This tensor aggregates information across multiple examples, allowing for broader statistical analyses. Users can apply summarization techniques such as mean, median, or any aggregating function to identify consistent patterns in the behavior of the model. These insights can be presented using heatmaps, clustering diagrams, or other structured visualizations.

\subsection{Case Study Setup}

To answer our research question \ref{rq:applicability} about the model's overall behavior, we use the global interpretability tensor ($\Phi$) to conduct three distinct statistical analyses, which directly correspond to the results presented in Section~\ref{sec:exploration}.

\subsubsection*{\textbf{\ref{rq:applicability_one}}- \textbf{Rational Probability Analysis.}} 

Our goal with this analysis was to justify the need for aggregation quantitatively. We hypothesized that raw $L_1$ tokens are too noisy for interpretation, and that $L_2$ concepts are necessary. Our probability analysis proceeds in two stages. 

First, we extracted the token-level probabilities and mapped them to the \levelone~concept level. For each sequence, we compute 30 concept-aligned rationale matrices \( \phi_\mathcal{C} \) and aggregate them into a third-order tensor \( \mathbb{T} \in \mathbb{R}^{|\mathcal{C}| \times |\mathcal{C}| \times 30} \). For each concept pair \( (c_i, c_j') \), we summarize the rationale probabilities across trials using medians and 75th percentiles, and construct bootstrapped confidence intervals to generate global explanations. This summarization reduces error propagation via the central limit theorem and enables aggregation into higher-level concepts, as the weak token-level signal motivates a more abstract and broadly distributed representation.

Next, at \leveltwo, we investigate the typical influence strength of our abstract categories. For each programming category, we compute the median of all its constituent token probabilities gathered from across all sequences and 30 trials. This allows us to determine the category's overall central tendency.

However, even after such aggregation, one critical question remains: are these observed patterns truly meaningful, or could they still arise from random noise? While median-based summaries help reduce stochasticity, they do not by themselves confirm whether the extracted rationale distributions are semantically structured. To address this concern and quantitatively validate the order within these probabilistic patterns, we complement our probability analysis with an information-theoretic perspective.

\textbf{Entropy Analysis.} To formally quantify the uncertainty of the rationale distributions, we conduct an information-theoretic analysis using Shannon Entropy\cite{shanonEntropy}. This approach is well-founded, as it is a standard practice to frame noise reduction as an entropy-minimization problem; a structured, \textit{clean} signal is inherently more ordered and has lower entropy than \textit{noisy}, random data~\cite{wiggins1978minimum}. In XAI, lower-entropy explanations are similarly considered more concise and informative~\cite{chen2018learning}, and information theory is used to validate that a data grouping (\ie our categories) is semantically meaningful rather than random~\cite{meila2007comparing}.

We compute the entropy for two distinct probability distributions, both derived from the pooled rationale data of our testbed: (1) The \textbf{\textit{$L_1$} (Token) Distribution}, based on the pooled frequencies of all raw rationale \textit{tokens}, and (2) The \textbf{\textit{$L_2$} (Concept) Distribution}, based on the pooled frequencies of our aggregated AST-based \textit{categories}. We calculate entropy in bits as $H(\mathbf{p}) = -\sum_i p_i\,\log_2 p_i$. This allows for a direct comparison of the uncertainty at each level of abstraction, with the results presented in Sec.~\ref{sec:entropy-results}.

\subsubsection*{\textbf{\textbf{\ref{rq:applicability_two}-\leveltwo  Rationale Frequency Analysis}}}
To determine which categories the model relies on most often, we count the number of times tokens belonging to each category appear as rationales. This count is performed for each of the 30 trials, resulting in a distribution of frequency counts for each category. We then compute descriptive statistics for this distribution (\ie mean, median, standard deviation), and a bootstrapped confidence interval, to robustly measure how frequently the model uses each category.

\subsubsection*{\textbf{\ref{rq:applicability_global}-\textbf{\leveltwo Conceptual Dependency Analysis}}}
To understand which types of concepts influence the generation of other concepts, we analyze the global tensor \( \Phi \). While this tensor can be visualized as a full heatmap, presenting such a large matrix directly can overwhelm the reader and obscure key patterns. Therefore, to improve clarity and focus on the most significant trends, we present the data in a series of abstracted data tables (see \tabref{tab:rationales-probability}).

Table \ref{tab:rationales-probability} displays a representative $10x10$ subset of the most relevant source and target categories. Each cell \((i, j)\) in a table shows the aggregated strength of influence (\eg the median probability) from source category \(i\) to target category \(j\) for a specific testbed. This format allows for a clear comparison of broad dependency patterns, such as a model's reliance on syntactic versus semantic information, across different experimental conditions. The complete data, corresponding to the full heatmaps, is available in our online appendix \cite{anonymized_repo} for further exploration.
\end{enumerate}

\subsubsection*{\textbf{\ref{rq:applicability_three} Encoder-Decoder \& Decoder-Decoder Interpretability}}\label{sec:e-d-interpretability}

To demonstrate the generalizability of \codeRational, we used the encoder-decoder model \bart for test generation via Fairseq\cite{ott2019fairseqfastextensibletoolkit}. We fine-tuned \bart with the \texttt{Methods2Test} dataset using BPE encoding. In this setting, the encoder processes a \textit{source} code context window (\eg \class, \focalmethod, \signature), while the decoder generates corresponding test code as the \textit{target} output.

After fine-tuning the \bart model, we sampled the model using a maximum of 150 generated tokens to extract and group the rationales. The maximum number of generated tokens proved sufficient, as we observed an average of 83 and a third quartile of 107 tokens per generated test. The sampling process extracts the rationales in \levelone as described in \ref{step2} on 1K test generation samples. To validate our experiment, we performed 30 sampling trials to obtain $\Vec{\phi_\mathcal{C}}$ following \ref{step3}. Next, we execute \ref{step4} to compute the $\Phi$ tensor. Using an encoder-decoder model, we also explored computing the rationales for the decoder component, in other words, using the same target test case with the decoder component as an autoregressive model. Finally, to visualize these rationales, we create a set of heatmaps based on the $\Phi$  tensor.



%% file: text/6.User_study.tex
\section{\codeRational in Practical Settings}
\label{sec:user_study}

To evaluate the usability of \codeRational in practical settings (\ref{rq:utility}), we designed a user study targeting practitioners with experience in ML and software engineering. We followed a purpose-sampling approach~\cite{baltes2021sampling}, reaching out to 86 experts in academia and industry, including students, researchers, developers, and data scientists who use \lmc for downstream SE tasks. After filtering out incomplete or low-quality responses, we retained 37 valid submissions. The study was hosted on Qualtrics~\cite{noauthor_qualtrics_nodate}. For detailed information on our participants, please refer to ~\tabref{tab:demograhics}.

\subsection{Survey Structure}

The survey was divided into four sections\footnote{See Appendix and \cite{anonymized_repo} for the full survey.}. The \textbf{first section} collected participants' backgrounds in Python, Java, and ML-based code generation, including years of experience and use cases. The \textbf{second section} showed 10 model-generated examples (5 Python completions, 5 Java test generations) and asked participants to identify responsible tokens for selected outputs and assess the correctness of each prediction. In the \textbf{third section}, participants reviewed \codeRational visualizations for the same examples and rated their agreement with the generated rationales using a 5-point Likert scale. Finally, in the \textbf{fourth section} , participants were asked to reflect on the general usefulness and interpretability of the framework.

\subsection{Qualitative Metrics}
\label{survey_metric}
We assess the usability of \codeRational using four qualitative dimensions:

\begin{enumerate}[label= \textbf{[m$_{\arabic*}$]:}, ref=m$_{\arabic*}$, wide,labelindent=5pt]\setlength{\itemsep}{0.2em}
\item \textbf{Usefulness.} 
\label{user_study_usefulness} Measures perceived practical value of code-based explanations in SE tasks~\cite{linardatosExplainable, management_solutions_xai, trustincollab}. The participants rated the usefulness based on example explanations and also provided open feedback.

\item \textbf{Readability.} 
\label{user_study_readability}
Assesses how informative and comprehensible the visualizations are, aligned with the desideratum informativeness. We asked participants to evaluate both the AST and the context-level visual formats.


\item \textbf{Alignment.} 
\label{user_study_alignment}
Measures how closely the generated rationale reflects human reasoning. The participants manually reported rationales, which we compared with the output of \codeRational using the Jaccard similarity. A higher score indicates stronger agreement.
\end{enumerate}

\subsection{Sample Selection}

We selected 10 diverse examples (5 per task), combining 2 correct predictions and 8 incorrect predictions containing syntactic and semantic errors documented in previous work~\cite{synerror,synerror1,synerror2}. Two authors independently labeled the model outputs using known error taxonomies; a third author validated the labels to ensure objectivity. Including flawed and correct samples helped to evaluate how \codeRational performs in various scenarios.
\input{tabs/survey_details}


\subsection{Validity}

To improve instrument validity, we conducted a pilot study with five external participants, which led to refinements in the phrasing of the questions and the diversity of error cases. We also consulted with an expert on qualitative methods. Two authors independently coded and analyzed open responses; disagreements were resolved through discussion to ensure interpretive consistency.

%% file: tabs/survey_details.tex
\begin{table}[]
\centering
\caption{Respondent Demographics on 37 participants}
\resizebox{\linewidth}{!}{%
\begin{tabular}{lllrllllr}
\multicolumn{1}{c}{\textbf{Attribute}} & \multicolumn{1}{c}{\textbf{Detail / Level}} & \textbf{N} & \multicolumn{1}{c}{\textbf{\%}} & \multicolumn{1}{c}{\textbf{}} & \multicolumn{1}{c}{\textbf{Attribute}} & \multicolumn{1}{c}{\textbf{Detail / Level}} & \textbf{N} & \textbf{\%} \\ \cline{1-4} \cline{6-9} 
\multirow{2}{*}{\textit{\textbf{\begin{tabular}[c]{@{}l@{}}Participant\\  Role\end{tabular}}}} & Researcher & 15 & 41\% &  & \multirow{3}{*}{\textit{\textbf{\begin{tabular}[c]{@{}l@{}}Python \\ Experience\end{tabular}}}} & Beginner & 3 & 8\% \\
 & Student & 22 & 59\% &  &  & Intermediate & 24 & 65\% \\ \cline{1-4}
\multirow{8}{*}{\textit{\textbf{\begin{tabular}[c]{@{}l@{}}Primary \\ Use Case*\end{tabular}}}} & Test Case Generation & 8 & 9\% &  &  & Expert & 10 & 27\% \\ \cline{6-9} 
 & Bug Fixing \& Program Repair & 17 & 19\% &  & \multirow{3}{*}{\textit{\textbf{\begin{tabular}[c]{@{}l@{}}Java \\ Experience\end{tabular}}}} & Beginner & 8 & 22\% \\
 & Program Synthesis & 8 & 9\% &  &  & Intermediate & 25 & 68\% \\
 & Refactoring & 7 & 8\% &  &  & Expert & 4 & 11\% \\ \cline{6-9} 
 & Program Translation & 4 & 4\% &  & \multirow{4}{*}{\textit{\textbf{\begin{tabular}[c]{@{}l@{}}SE \\ Experience**\end{tabular}}}} & 0 to 3 years & 16 & 44\% \\
 & Code Completion & 16 & 18\% &  &  & 3 to 7 years & 10 & 28\% \\
 & Code Generation & 21 & 23\% &  &  & 7 to 15 years & 7 & 19\% \\
 & Other & 10 & 11\% &  &  & 16+ years & 3 & 8\% \\ \bottomrule
\end{tabular}%
}
\tiny *\textit{The primary use case is a multiple-choice question.}
\tiny ** \textit{One participant did not disclose the experience.} 

\label{tab:demograhics}
\end{table}

%% file: text/6_rq1_result.tex
\section{\codeRational Exploratory Analysis}\label{sec:exploration}

The exploratory analysis aims to evaluate the applicability of \codeRational and the answer \ref{rq:applicability}. We introduce an exploratory methodology, enumerate the results, and discuss our findings on applying \codeRational.

\subsection{Code Generation Interpretability}
\subsubsection{Rationale Probability Distribution Analysis}
\label{result:distribution}
\input{tabs/7_Research_questions/rationales_frequency_TB2}

Analysis in \textbf{ \levelone} consistently revealed extremely low probability scores. The medians for these token-level rationales typically hovered below 0.012, with the 75th percentile hovering around 0.066. This indicates a highly skewed distribution in which the vast majority of individual tokens, when considered in isolation, contribute minimally to the model's prediction. 

However, a different pattern emerged when we analyzed the distribution at \textbf{\leveltwo}. As shown in \tabref{tab:rational_frequency} (column ``RQ1.1 Median Probability''), after aggregating the token probabilities into their respective categories, the median scores are significantly higher clustering around 0.06-0.07 for most programming languages. This represents a roughly \textbf{6-fold increase} in probability compared to the median at the token level. This amplification demonstrates that while the influence of any single token is weak, the collective influence of tokens belonging to a meaningful category is substantial.

\subsubsection{Entropy Analysis Results}
\label{sec:entropy-results}

The results of this analysis, presented in Table~\ref{tab:entropy-validation}, provide strong quantitative validation for our aggregation approach. We find that the raw $L_1$ token distribution is high-entropy, with values ranging from 18.18 to 20.89 bits across our four testbeds.

In stark contrast, after aggregating these tokens into our $L_2$ AST-based categories, the entropy drops dramatically to a lower value of approximately 9.9 bits for all testbeds. This represents a significant reduction in entropy ($\Delta H$) of -8.44 to -10.91 bits($\gtrsim 50\%$). This drop confirms that the strong signal amplification reported in Sect.~\ref{result:distribution} is not a statistical artifact of aggregation. Rather, it provides information-theoretic evidence that our conceptual mapping successfully distills a predictable, low-entropy signal from the high-entropy raw token data, thereby capturing genuine, structured reasoning patterns.

\input{tabs/7_Research_questions/entropy}



\textbf{Discussion.} The finding of overwhelmingly low probabilities at \levelone suggests that the model's reasoning is not driven by a few \textit{highly influential} tokens. Instead, its decisions seem to arise from a broad, distributed attention to the overall context. This presents a significant challenge: \levelone explanations, based on single tokens, are not only inherently weak but also statistically noisy, as confirmed by our high-entropy finding. They are also misaligned with a developer's mental model, which is confused by tokenization (e.g., splitting \texttt{readFile()} into \texttt{`read'}, \texttt{`File'}, etc.).

Our framework addresses this by grouping sub-tokens into coherent units (\ie \textbf{identifier}) and aggregating their probabilities. This significantly amplifies the signal (observed at \leveltwo) and is validated by our information-theoretic analysis, which confirms a >50\% drop in entropy. This key result of collapsing low-signal, high-noise tokens into meaningful categories enables a statistically valid summary of the model’s central tendencies. This methodological shift aligns explanations with developer reasoning (concepts vs. tokens), making them more understandable and resilient to noise.

\begin{boxK}
\ref{rq:applicability_one}: Our analysis shows \texttt{L1} rationales are weak (median < 0.012) and noisy (high-entropy). Aggregating them into \texttt{L2} amplifies the signal $\approx$6x (median $\approx$ 0.07) and reduces entropy by >50\%, justifying our global, concept-level analysis.

\end{boxK}

\subsubsection{\leveltwo Rationale Frequency Analysis}
The analysis of the frequencies of the rationale category at the \leveltwo level, summarized in Table~\ref{tab:rational_frequency}, reveals a clear and significant hierarchy in the types of categories the model uses for its predictions. Some structural and syntactic categories overwhelmingly dominate the model's reasoning process.

Within the Programming Language taxonomy, \structural concepts serve as the most common rationale by a large margin, appearing on average $\approx 1.19$ million times across the trials. This is followed by other general categories like \statements ($\approx 469k$) and the non-semantic category \expression ($\approx 341k$). Among categories related to Natural Language, \texttt{[nl\_noun]} is also highly frequent ($\approx 360k$). In stark contrast, categories that represent specific, high-level programming logic appear orders of magnitude less often. For example, \asserts ($\approx 1k$), \bool ($\approx 1.7k$), and \loops ($\approx 1.7k$) are among the least frequent rationale categories, indicating that the model consults them far less often when making a prediction.

\textbf{Discussion.} This frequency analysis provides a powerful global view of the model's reasoning strategy, revealing a strong preference for syntactic heuristics over semantic understanding. The overwhelming frequency of categories such as \structural, \statements, and \expression suggests that the model has learned to rely on shallow but reliable syntactic and structural patterns to predict the next token. For a generative task, this is a logical strategy, as adhering to local syntax is a robust way to achieve correctness.

However, this behavior diverges significantly from how a human developer thinks. A developer prioritizes the semantics of \loops or \conditionals to understand and formulate the program logic. However, our results show that the model consults these specific and semantically rich categories much less often than it consults general syntax. This reliance on syntactic shortcuts, a form of model overinterpretation, directly impacts trustworthiness. It suggests a model that excels at pattern-matching the surface structure of code but may lack a deep functional understanding of the logic it generates. A developer, therefore, cannot assume that the model's reasoning process mirrors their own and must be cautious of outputs that are syntactically plausible but logically flawed.

\begin{boxK}
{\ref{rq:applicability_two}: Our rationale frequency analysis reveals a global preference for shallow syntactic heuristics. The model relies heavily on concepts like \structural and \expression while using semantically rich categories such as \loops and \conditionals far less often. This pattern suggests a strong surface-level pattern matching but a limited understanding of program logic.
}
\end{boxK}

\subsubsection{Level 1 Conceptual Dependency Analysis}
Our analysis reveals a substantial shift in the model’s conceptual reasoning strategy as the prompt evolves from structurally rich to semantically rich contexts as shown in \ref{tab:rationales-probability}. In the \sgbd testbed, where code dominates the prompt, the model exhibits a strong reliance on syntactic features and marked self-dependence. The \asserts category shows the highest self-influence (0.410), and dependencies such as \exceptions influencing \punctuation (0.647) suggest that the model draws on structural heuristics, likely reflecting learned code templates. In contrast, natural language concepts exert limited influence on code prediction; for example, the impact of \nlnoun on \return remains low (0.113).

As the input context expands in \dcsgbd, the model begins to incorporate semantically meaningful patterns. While \texttt{asserts} remains largely self-reliant (0.327), notable cross-category dependencies emerge, such as \functional influencing \asserts (0.289). This indicates a transition from shallow pattern matching to reasoning informed by higher-level functional semantics.

In \dc, where only descriptions in natural language guide the generation process, the behavior of the model changes markedly. The strongest dependency now appears between \asserts and \exceptions (0.490), a high-level semantic link absent in more code-centric settings. Additionally, the influence of natural language tokens becomes more pronounced, for instance, \nlnoun substantially affects the generation of \oop (0.185) and \bool (0.167). Across all testbeds, it is also notable that the punctuation category consistently maintains a low but non-zero influence on most generated code categories.
\input{tabs/7_Research_questions/rationale-code-generation}

\textbf{Discussion.}
This analysis reveals the remarkable adaptability of the model and underscores the complexity of its reasoning process. Rather than applying a fixed set of rules, the model dynamically adjusts its reasoning strategy based on available prompt information.

Concrete shifts in dependency patterns illustrate this adaptability. In code-dominant settings, strong self-reliance, evident in categories such as \asserts, suggests that the model has learned an effective heuristic. Similarly, the presence of non-zero punctuation categories across testbeds indicates a form of overinterpretation, where the model relies on low-information features as a fundamental heuristic. On the other hand, the emergence of dependencies from docstring nouns and verbs to identifiers and statements signals a deeper capacity to map natural language semantics to code structure.

This is further reflected in the evolving behavior of the \loops category, whose dependencies shift from structural elements (\eg \return, \oop) to logical constructs (\eg \conditionals, \bool) and eventually to validation constructs (\asserts) as the prompt becomes more semantically complete.

The \dc setting exemplifies the adaptive behavior of the model. Despite the absence of code in the initial prompt, the model uses its own autoregressive output as contextual signals. As it generates early code tokens, it begins to condition on them, effectively bootstrapping its reasoning from natural language to code syntax mid-generation.

\begin{boxK}
\ref{rq:applicability_global}: Our global, concept-level analysis reveals dynamic shifts in reasoning, self-reliance, and movement from syntactic heuristics to semantically informed patterns as context changes. These high-level dependencies remain hidden in token-level or accuracy-based evaluations. Our method enables the uncovering of such model behaviors.
\end{boxK}


\subsection{Test Case Generation Interpretability}
The test generation analysis comprises the application of \codeRational using the \bart encoder-decoder model (see \secref{sec:e-d-interpretability}). Each model sampling generates the rationale from the source and target (\ie source code and test case). We observe that each sampling iteration reduces the number of generated tokens, starting from an average of 83 tokens to 46.55 tokens in length by the 30th trial. 

\begin{figure}[t]
  \centering
  \vspace{-0.2cm}
  
  \includegraphics[width=\linewidth]{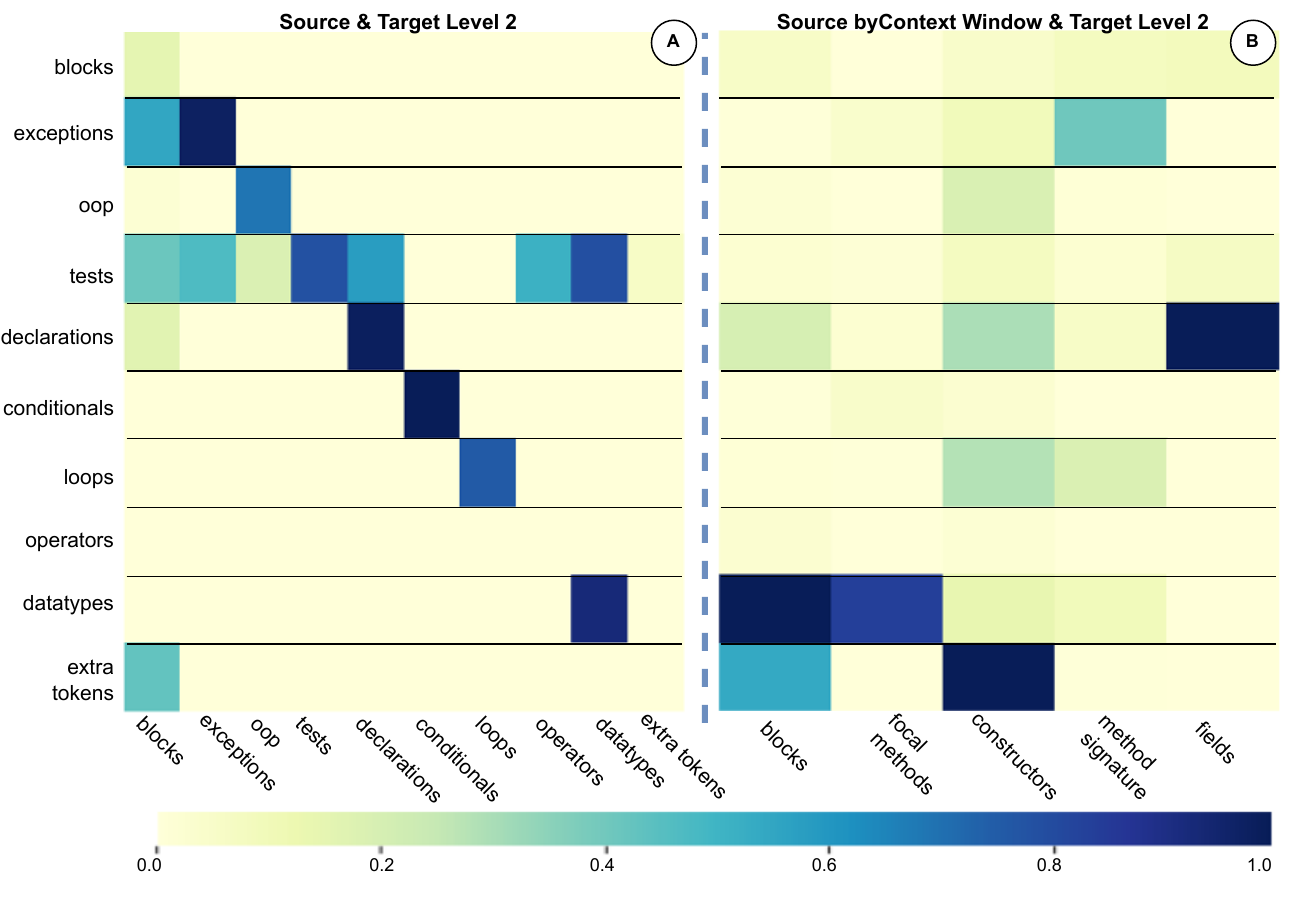}
  
  \vspace{-0.3cm}
  
  \caption{Category and Context window level rationales}
  \label{fig:context_level}
  
  \Description{The figure contains two heatmaps (A and B) visualizing rationale probabilities. Heatmap A (Source & Target Level 2) shows a strong diagonal pattern, indicating self-reliance where source concepts largely influence identical target concepts. Heatmap B (Context Window) maps components like 'focal method' and 'fields' to target concepts, showing for instance that focal methods strongly influence the generation of data types.}
  
  \vspace{-0.5cm}
\end{figure}

\figref{fig:context_level} depicts to global results using the encoder-decoder architecture.  The  \figref{fig:context_level}-\circled{A} shows a heatmap with the source to target rationales. The strong diagonal pattern indicates consistent \textbf{self-reliance} between the source and target concepts. However, we observe weaker associations for concepts such as \texttt{[blocks], [operators], and [extra\_tokens]}, suggesting low influence on the generation of their corresponding target. 
Interestingly, while \texttt{[operators]} in the target are often generated from the same concept in the source, other target concepts, such as \texttt{[test]} which includes ``assert'' statements for test generation, are influenced mainly by a broader range of input concepts including \texttt{[data types], [test], [operators],} and \texttt{[conditionals]}. This indicates that certain output concepts are shaped by a combination of multiple input-level abstractions.

\figref{fig:context_level}-\circled{B} shows a second heatmap that groups the input source by the \textit{context window} (\ie \texttt{[fields],[signature], [constructor]}) (see \figref{fig:ewash_taxonomy}). In this analysis, we report only the minimum rational probability, as the average illustrates the influence of multiple concepts on each other. We observe that  \texttt{[blocks]} and  \texttt{[focal methods]} mostly impact the  \texttt{[data types]} generation. At the same time, the field influences the generation of  \texttt{[declarations]} for keywords in Java like \texttt{``void'', ``static''}, or \texttt{``import''}. 

The target-to-target analysis using the decoder component of \bart model (\ie an auto-regressive architecture) indicates only the influence of the \textit{[oop]} concepts on the \textit{[declaration]} output with a probability of 0.9, while the other concepts are close to zero (see appendix\cite{anonymized_repo}).

From our data analysis, we computed the average influence of the context window on token prediction for a test generation. As expected, the most influential context window is the \textit{focal method} with an average probability impact of 0.5, followed by the method signature with 0.22 and class with 0.14. However, we observe the field context with the lower probability of 0.006 but the highest impact with $\approx 11K$ tokens.

\textbf{Discussion.} The test case generation demonstrates the applicability of \codeRational on encoder-decoder transformer models. The model focuses attention on specific concepts, such as data types or operators, to generate assertions in a Java unit test, rather than relying on loops and conditionals. \codeRational helps confirm the high relevance of focal methods in the generation of tests. \codeRational enabled the analysis of the decoder component on an encoder-decoder architecture; however, the results report a high correlation between the \texttt{oop} and \texttt{declarations}. The use of heatmaps improves the explanation of what the model uses to link source-target in SE terms.
\begin{boxK}
\ref{rq:applicability_three}: \codeRational performs effectively in encoder-decoder architectures, such as \bart, generating focused and interpretable rationales that align accurately with the source code and test cases.

\end{boxK}

\subsection{Threats to validity}


Although our analysis centers on mid-scale models (\texttt{codeparrot\allowbreak-small} and \texttt{BART-large}), this choice follows directly from the methodological requirements of our framework. The approach relies on \textbf{white-box access} for fine-tuning to ensure compatibility with our setup (\secref{sec:int_setup}), and the multi-trial evaluation introduces substantial \textbf{computational demands}. Within these constraints, mid-scale models offered a practical setting for validating the method. The focus of the work is the introduction of a \textbf{transformer-agnostic interpretability framework}, and the selected models function as representative platforms rather than objects of comparison. The underlying technique, \codeRational, applies to any autoregressive transformer with a final softmax output layer~\cite{vaswani_attention_2017}. As the community releases larger pretrained models with accessible weights, particularly those optimized for safety and alignment~\cite{bai2022constitutionalaiharmlessnessai}, the range of compatible targets will expand to include modern open checkpoints. The framework is therefore positioned to scale as model accessibility grows, supporting broader empirical validation. This approach is common in interpretability research, where established models are used to study fundamental behaviors \cite{morris2025languagemodelsmemorize}, and it aligns with empirical SE norms~\cite{wohlin2012experimentation}, where generalization strengthens through the accumulation of complementary studies.

Second, our mapping from tokens to programming categories is a form of abstraction. Although our taxonomy is grounded in standard AST constructs, we acknowledge that other categorization schemes are possible. The specific categories chosen could influence the results. However, we argue that some form of abstraction is a necessary step to move from machine-level tokens to human-centric explanations, and we provide our full taxonomy in the online appendix for transparency.

%% file: tabs/7_Research_questions/rationales_frequency_TB2.tex
\begin{table}[]
\caption{Rational probability and frequency on  \sgbd}
\centering
\resizebox{\linewidth}{!}{%
\begin{tabular}{llcclrrrrr}
\multicolumn{2}{c}{\textbf{Taxonomy}} & \textbf{} & \multicolumn{1}{c}{\ref{rq:applicability_one} \textbf{Probability}} &  & \multicolumn{5}{c}{\textbf{\ref{rq:applicability_two}: Ranking by Frequency}} \\ \cline{1-2} \cline{4-4} \cline{6-10} 
\multicolumn{1}{c}{\textbf{Group \levelthree}} & \multicolumn{1}{c}{\textbf{Category \leveltwo}} & \textbf{} & \multicolumn{1}{c}{\textbf{Median}} &  & \multicolumn{1}{c}{\textbf{Mean}} & \multicolumn{1}{c}{\textbf{STD}} & \multicolumn{1}{c}{\textbf{Median}} & \multicolumn{1}{c}{\textbf{CI -Low}} & \multicolumn{1}{c}{\textbf{CI -High}} \\ \hline
\multicolumn{10}{c}{\textit{\textbf{Programming Language Modality}}} \\ \hline
 & \textit{structural} &  & 6.42E-02 &  & \cellcolor[HTML]{CBCEFB} {\ul 6.91E+04} & 4.09E+03 & 6.87E+04 & 6.76E+04 & 7.06E+04 \\
 & \textit{statements} &  & 6.62E-02 &  & 2.65E+04 & 4.75E+03 & 2.68E+04 & 2.47E+04 & 2.82E+04 \\
 & \textit{types} &  & 6.19E-02 &  & 8.42E+03 & 1.83E+03 & 8.43E+03 & 7.74E+03 & 9.10E+03 \\
 & \textit{with} &  & 6.22E-02 &  & 5.73E+03 & 1.38E+03 & 5.54E+03 & 5.22E+03 & 6.23E+03 \\
 & \textit{oop} &  & \cellcolor[HTML]{CBCEFB}{\ul 1.04E-01} &  & 6.98E+02 & 9.85E+01 & 6.97E+02 & 6.61E+02 & 7.34E+02 \\
 & \textit{conditionals} &  & 7.04E-02 &  & 3.28E+02 & 8.64E+01 & 3.15E+02 & 2.96E+02 & 3.60E+02 \\
 & \textit{loops} &  & 9.24E-02 &  & \cellcolor[HTML]{FFCCC9} {\ul 2.61E+02} & 7.87E+01 & 2.56E+02 & 2.32E+02 & 2.91E+02 \\
 & \textit{asserts} &  & \cellcolor[HTML]{FFCCC9}{\ul 3.57E-02} &  & 1.62E+02 & 3.89E+01 & 1.53E+02 & 1.47E+02 & 1.76E+02 \\
 & \textit{bool} &  & \cellcolor[HTML]{CBCEFB}{\ul 1.01E-01} &  & \cellcolor[HTML]{FFCCC9}{\ul 1.12E+02} & 5.36E+01 & 1.04E+02 & 9.22E+01 & 1.32E+02 \\
\multirow{-10}{*}{\textit{\textbf{Semantic}}} & \textit{exceptions} &  & \cellcolor[HTML]{CBCEFB}{\ul 1.87E-01} &  & 4.28E+01 & 5.23E+01 & 1.60E+01 & 2.34E+01 & 6.22E+01 \\ \hline
 & \textit{identifiers} &  & 6.48E-02 &  & 6.93E+04 & 2.11E+03 & 6.93E+04 & 6.85E+04 & 7.01E+04 \\
 & \textit{string} &  & 5.15E-02 &  & 6.74E+03 & 8.34E+02 & 6.82E+03 & 6.44E+03 & 7.05E+03 \\
\multirow{-3}{*}{\textit{\textbf{\begin{tabular}[c]{@{}l@{}}Natural\\ Language \\ in Code\end{tabular}}}} & \textit{comments} &  & 6.21E-02 &  & 3.48E+03 & 7.57E+02 & 3.42E+03 & 3.20E+03 & 3.76E+03 \\ \hline
\textit{\textbf{Error}} & \textit{errors} &  & 6.28E-02 &  & 1.80E+05 & 6.62E+03 & 1.80E+05 & 1.77E+05 & 1.82E+05 \\ \hline
 & \textit{expression} &  & 6.00E-02 &  & \cellcolor[HTML]{CBCEFB} {\ul 1.17E+05} & 2.26E+04 & 1.14E+05 & 1.09E+05 & 1.26E+05 \\
 & \textit{punctuation} &  & 5.78E-02 &  & 3.66E+04 & 1.06E+03 & 3.64E+04 & 3.62E+04 & 3.70E+04 \\
 & \textit{operators} &  & 6.63E-02 &  & 5.34E+03 & 7.35E+02 & 5.24E+03 & 5.07E+03 & 5.61E+03 \\
 & \textit{return} &  & 6.45E-02 &  & 1.66E+03 & 1.08E+02 & 1.65E+03 & 1.62E+03 & 1.70E+03 \\
\multirow{-5}{*}{\textit{\textbf{\begin{tabular}[c]{@{}l@{}}Non\\ Semantic\end{tabular}}}} & \textit{functional} &  & 8.40E-02 &  & 3.79E+02 & 2.03E+02 & 3.26E+02 & 3.04E+02 & 4.54E+02 \\ \hline
\multicolumn{10}{c}{\textit{\textbf{Natural Language Modality}}} \\ \hline
 & \textit{noun} &  & 6.53E-02 &  & 6.82E+04 & 2.17E+03 & 6.76E+04 & 6.74E+04 & 6.90E+04 \\
 & \textit{verb} &  & 5.25E-02 &  & 3.80E+03 & 3.99E+02 & 3.80E+03 & 3.65E+03 & 3.95E+03 \\
 & \textit{adjetive} &  & 5.32E-02 &  & 1.66E+03 & 4.36E+02 & 1.67E+03 & 1.50E+03 & 1.82E+03 \\
\multirow{-4}{*}{\textit{\textbf{Semantic}}} & \textit{pronoun} &  & 4.80E-02 &  & 6.57E+01 & 6.54E+01 & 4.30E+01 & 4.15E+01 & 8.99E+01 \\ \hline
 & \textit{adverb} &  & 5.41E-02 &  & 7.14E+02 & 1.90E+02 & 6.82E+02 & 6.44E+02 & 7.85E+02 \\
 & \textit{determier} &  & 4.02E-02 &  & 6.26E+02 & 1.64E+02 & 6.28E+02 & 5.66E+02 & 6.87E+02 \\
 & \textit{preposition} &  & 4.79E-02 &  & 5.66E+02 & 1.61E+02 & 5.45E+02 & 5.07E+02 & 6.26E+02 \\
 & \textit{cardinal} &  & 5.68E-02 &  & 2.02E+02 & 1.68E+02 & 1.58E+02 & 1.40E+02 & 2.64E+02 \\
 & \textit{modal} &  & 6.64E-02 &  & 5.19E+01 & 5.20E+01 & 4.50E+01 & 3.27E+01 & 7.12E+01 \\
 & \textit{conjunction} &  & \cellcolor[HTML]{FFCCC9}{\ul 2.52E-02} &  & 3.67E+01 & 3.46E+01 & 3.15E+01 & 2.39E+01 & 4.95E+01 \\
\multirow{-7}{*}{\textit{\textbf{\begin{tabular}[c]{@{}l@{}}Non\\ Semantic\end{tabular}}}} & \textit{particle} &  & \cellcolor[HTML]{FFCCC9}{\ul 1.03E-02} &  & 3.40E+00 & 8.16E+00 & 0.00E+00 & 3.81E-01 & 6.42E+00 \\ \bottomrule
\end{tabular}%
}
\tiny \textit{Red indicates the lowest, and blue the highest, rationale probability observed for each concept pair.}
\label{tab:rational_frequency}
\end{table}

%% file: tabs/7_Research_questions/entropy.tex
\begin{table}[t]
  \centering
  \vspace{-0.2cm}
  \footnotesize
  \setlength{\tabcolsep}{8pt}
  \caption{Shannon Entropy (in bits) of Pooled Rationale Distributions at the Token ($L_1$) and Concept ($L_2$) levels.}
  \label{tab:entropy-validation}
  \vspace{-0.3cm}
  \resizebox{\linewidth}{!}{%
  \begin{tabular}{@{}lrrrr@{}}
    \textbf{Testbed} & \textbf{$H(L_1)$} & \textbf{$H(L_2)$} & \textbf{$\Delta H$} & \textbf{\% Change} \\
    \midrule
    $TB_1$ (SG\_BD) & 18.18 & 9.74 & -8.44 & -46.4\% \\
    $TB_2$ (DC\_SG\_BD) & 20.83 & 9.95 & -10.88 & -52.2\% \\
    $TB_3$ (DC\_SG) & 20.89 & 9.98 & -10.91 & -52.2\% \\
    $TB_4$ (DC) & 19.82 & 9.93 & -9.89 & -49.9\% \\
    \bottomrule
  \end{tabular}}
  \tiny \textit{The \% Change column shows the relative reduction of entropy in $L_2$ compared to $L_1$.}
  \vspace{-0.8cm}
\end{table}

%% file: tabs/7_Research_questions/rationale-code-generation.tex
\begin{table}[]
\centering
\caption{Mean probability at \leveltwo on each testbed}
\resizebox{\linewidth}{!}{%
\begin{tabular}{lllllllllll}

 & \textbf{asserts} & \textbf{oop} & \textbf{loops} & \textbf{funct.} & \textbf{excep.} & \textbf{bool} & \textbf{punct.} & \textbf{nl\_par.} & \textbf{nl\_noun} & \textbf{return} \\ \hline
 \multicolumn{11}{c}{\textit{\textbf{Rationales  probability \leveltwo for  \sgbd}}} \\ \hline
\textit{\textbf{asserts}} & \cellcolor[HTML]{CBCEFB}{\ul 0.410} & 0.242 & 0.000 & 0.000 & 0.000 & 0.038 & 0.070 & 0.000 & \cellcolor[HTML]{FFCCC9}0.019 & 0.000 \\
\textit{\textbf{oop}} & 0.000 & 0.184 & 0.066 & 0.013 & 0.040 & 0.201 & 0.261 & 0.000 & 0.054 & 0.157 \\
\textit{\textbf{loops}} & 0.000 & 0.000 & 0.069 & 0.002 & 0.060 & 0.088 & 0.119 & 0.000 & 0.028 & 0.254 \\
\textit{\textbf{functional}} & 0.000 & 0.000 & 0.073 & 0.084 & 0.000 & 0.046 & 0.121 & 0.000 & 0.046 & 0.152 \\
\textit{\textbf{exceptions}} & 0.000 & 0.136 & 0.000 & 0.000 & 0.133 & 0.000 & \cellcolor[HTML]{CBCEFB}{\ul 0.647} & 0.000 & 0.044 & 0.134 \\
\textit{\textbf{bool}} & 0.107 & 0.123 & 0.042 & 0.000 & 0.070 & 0.113 & 0.203 & 0.000 & 0.028 & 0.120 \\
\textit{\textbf{punctuation}} & 0.062 & 0.089 & 0.088 & 0.070 & 0.063 & 0.091 & 0.131 & 0.045 & 0.037 & 0.100 \\
\textit{\textbf{nl\_particle}} & 0.000 & 0.000 & 0.000 & 0.000 & 0.000 & 0.000 & 0.000 & 0.000 & \cellcolor[HTML]{FFCCC9}0.007 & 0.000 \\
\textit{\textbf{nl\_noun}} & \cellcolor[HTML]{FFCCC9}0.056 & 0.138 & 0.070 & 0.081 & 0.067 & 0.102 & 0.132 & 0.040 & 0.043 & \cellcolor[HTML]{FFCCC9}0.113 \\
\textit{\textbf{return}} & 0.000 & 0.147 & 0.036 & 0.065 & 0.044 & 0.050 & 0.144 & 0.000 & 0.037 & 0.140 \\ \hline
\multicolumn{11}{c}{\textit{\textbf{Rationales  by concept   \leveltwo for \dcsgbd}}} \\ \hline
\textit{\textbf{asserts}} & \cellcolor[HTML]{CBCEFB}{\ul 0.327} & 0.192 & 0.033 & 0.049 & 0.015 & 0.048 & 0.091 & 0.000 & 0.029 & 0.007 \\
\textit{\textbf{oop}} & 0.100 & 0.193 & 0.053 & 0.051 & 0.062 & 0.211 & 0.188 & 0.035 & 0.044 & 0.201 \\
\textit{\textbf{loops}} & 0.151 & 0.123 & 0.033 & 0.071 & 0.063 & 0.139 & 0.088 & 0.019 & 0.025 & 0.012 \\
\textit{\textbf{functional}} & \cellcolor[HTML]{CBCEFB}{\ul 0.289} & 0.095 & 0.046 & 0.061 & 0.076 & 0.066 & 0.093 & 0.000 & 0.037 & 0.218 \\
\textit{\textbf{exceptions}} & 0.167 & 0.207 & 0.073 & 0.053 & 0.060 & 0.139 & 0.143 & 0.024 & 0.033 & 0.133 \\
\textit{\textbf{bool}} & 0.000 & 0.238 & 0.106 & 0.061 & 0.069 & 0.132 & 0.105 & \cellcolor[HTML]{FFCCC9} 0.020 & 0.030 & 0.192 \\
\textit{\textbf{punctuation}} & 0.068 & 0.164 & 0.090 & 0.064 & 0.057 & 0.090 & 0.102 & 0.050 & 0.034 & 0.119 \\
\textit{\textbf{nl\_particle}} & 0.000 & 0.002 & 0.037 & 0.000 & 0.035 & 0.000 & 0.117 & 0.002 & 0.021 & 0.000 \\
\textit{\textbf{nl\_noun}} & 0.128 & 0.018 & 0.081 & 0.065 & 0.057 & 0.113 & 0.101 & 0.077 & 0.038 & 0.138 \\
\textit{\textbf{return}} & \cellcolor[HTML]{FFCCC9} 0.015 & 0.160 & 0.101 & 0.029 & 0.054 & 0.055 & 0.086 & 0.055 & 0.033 & 0.157 \\ \hline
\multicolumn{11}{c}{\textit{\textbf{Rationales  probability \leveltwo for  \dcsg}}} \\ \hline
\textit{\textbf{asserts}} & \cellcolor[HTML]{CBCEFB}{\ul 0.264} & 0.166 & 0.154 & 0.000 & 0.018 & 0.118 & 0.133 & 0.002 & 0.032 & 0.031 \\
\textit{\textbf{oop}} & 0.052 & 0.154 & 0.087 & 0.048 & 0.072 & 0.097 & \cellcolor[HTML]{CBCEFB}{\ul 0.203} & 0.057 & \cellcolor[HTML]{FFCCC9}0.040 & 0.119 \\
\textit{\textbf{loops}} & 0.000 & 0.037 & 0.072 & 0.067 & 0.044 & 0.124 & 0.119 & 0.061 & 0.028 & 0.000 \\
\textit{\textbf{functional}} & 0.000 & 0.060 & 0.137 & 0.088 & 0.106 & 0.046 & 0.111 & 0.000 & 0.040 & 0.081 \\
\textit{\textbf{exceptions}} & 0.000 & 0.093 & 0.088 & 0.063 & 0.076 & 0.184 & \cellcolor[HTML]{CBCEFB}{\ul 0.223} & 0.124 & 0.039 & 0.138 \\
\textit{\textbf{bool}} & \cellcolor[HTML]{CBCEFB}{\ul 0.208} & 0.199 & 0.038 & 0.105 & 0.076 & 0.125 & 0.125 & 0.016 & 0.031 & 0.133 \\
\textit{\textbf{punctuation}} & 0.106 & 0.117 & 0.084 & 0.063 & 0.067 & 0.093 & 0.110 & 0.071 & 0.034 & 0.123 \\
\textit{\textbf{nl\_particle}} & 0.000 & 0.001 & 0.000 & 0.028 & 0.069 & 0.000 & 0.062 & 0.037 & 0.021 & 0.000 \\
\textit{\textbf{nl\_noun}} & 0.108 & 0.013 & 0.077 & 0.068 & 0.066 & 0.111 & 0.109 & 0.056 & 0.037 & 0.115 \\
\textit{\textbf{return}} & 0.025 & 0.173 & 0.166 & 0.058 & 0.040 & 0.049 & 0.098 & 0.000 & 0.034 & 0.144 \\ \hline
\multicolumn{11}{c}{\textit{\textbf{Rationales  probability \leveltwo for  \dc}}} \\ \hline
\textit{\textbf{asserts}} & 0.297 & 0.130 & 0.025 & 0.000 & \cellcolor[HTML]{CBCEFB}{\ul 0.490} & 0.130 & 0.090 & 0.000 & \cellcolor[HTML]{FFCCC9}0.022 & 0.000 \\
\textit{\textbf{oop}} & 0.070 & 0.249 & 0.059 & 0.018 & 0.052 & 0.256 & 0.250 & 0.020 & 0.040 & 0.180 \\
\textit{\textbf{loops}} & 0.000 & 0.159 & 0.077 & 0.158 & 0.068 & 0.138 & 0.139 & 0.000 & 0.031 & 0.203 \\
\textit{\textbf{functional}} & 0.000 & 0.179 & 0.147 & 0.078 & 0.000 & 0.048 & 0.115 & 0.000 & 0.061 & 0.193 \\
\textit{\textbf{exceptions}} & 0.070 & 0.125 & 0.082 & 0.000 & 0.065 & 0.172 & 0.196 & 0.066 & 0.038 & 0.147 \\
\textit{\textbf{bool}} & 0.114 & 0.316 & 0.068 & 0.074 & 0.060 & 0.156 & 0.150 & 0.000 & 0.033 & 0.135 \\
\textit{\textbf{punctuation}} & 0.102 & 0.153 & 0.087 & 0.078 & 0.062 & 0.109 & 0.121 & 0.079 & 0.037 & 0.161 \\
\textit{\textbf{nl\_particle}} & 0.000 & 0.096 & 0.000 & 0.057 & 0.041 & 0.000 & 0.077 & 0.030 & \cellcolor[HTML]{FFCCC9}0.023 & 0.054 \\
\textit{\textbf{nl\_noun}} & 0.102 & \cellcolor[HTML]{CBCEFB}{\ul 0.185} & 0.072 & 0.068 & 0.058 & \cellcolor[HTML]{CBCEFB}{\ul 0.167} & 0.116 & 0.053 & \cellcolor[HTML]{FFCCC9}0.037 & 0.149 \\
\textit{\textbf{return}} & 0.000 & 0.166 & 0.194 & 0.045 & 0.051 & 0.058 & 0.103 & 0.026 & 0.034 & 0.158 \\ \hline

\end{tabular}%
}
\tiny Purple: the highest values reported per dataset. Red: The lowest non-zero rational values.
\label{tab:rationales-probability}
\vspace{-3em}
\end{table}

%% file: text/7_rq2_result.tex
\section{User Study Results \& Discussion}\label{sec:user_results}

The summarized responses are detailed in \tabref{tab:survery_metrics} with the metrics discussed in \ref{survey_metric}. We also present results from open-ended questions.

\subsection{Results}

\textit{Usefulness \ref{user_study_usefulness}:}  User study results indicate that practitioners found \codeRational highly useful in various model development activities. A significant 89\% of practitioners agreed that \codeRational could be useful for fine-tuning models, with no participants disagreeing, highlighting its potential to reveal challenging code concepts for model improvement. Similarly, 84\% agreed that it is valuable to interpret the model output, and 81\% found it useful to infer causal relationships between input and output.

In open-ended feedback, users particularly praised \codeRational's ability to demonstrate these causal links and aid in interpreting model behavior, seeing its potential for debugging and designing better models. One user stated: \textit{``Useful for identifying what input data leads to generated code''}. Similarly, another user quoted \textit{``I think it is useful to get a more detailed look at what the model is doing. It would be useful for debugging and understanding the model better.''}

\input{tabs/survey}

\textit{Readability \ref{user_study_readability}:} 78\% of the participants found that the AST-based code explanations are informative to explain code completion. Moreover, none of the participants disagreed with the informativeness of the AST-based explanations. Similarly, 79\% of the participants found context-level-based explanations informative to explain test-case generations. However, only 52\% found AST-based graphical representations easy to read and 59\% found context-level graphical representations easy to read. This explains why our proposed taxonomy for code-based explanations of code-completion and test-case generation is solid; however, there is an opportunity for HCI researchers to improve the readability of graphical representations.



\textit{Alignment \ref{user_study_alignment}:} Our analysis revealed significant disparities between model-generated rationales and human reasoning for code-related tasks. For code completion samples, the Jaccard similarity score was just 0.074, while for test-case generation samples, it was 0.306. These low scores indicate a substantial difference between the rationales provided by the model and those given by humans.

This gap in reasoning is further evidenced by the participant agreement rates with the generated rationales. Only 53.6\% of the participants agreed with the model's explanations for code completion tasks, while 62\% agreed with the rationales for test-case generation tasks. These figures underscore a misalignment between the model and human reasoning processes.

\begin{boxK}
\ref{rq:utility}: \codeRational enables interpretable model development by revealing causal relationships between input and output, supporting tasks such as fine-tuning and debugging. Participants found it informative and readable for code completion and test generation, and useful for comparing model rationales with human reasoning to support calibrated trust.
\end{boxK}


\subsection{Discussion}


Our user study provides the human-centered perspective that complements our quantitative findings. The clearest insight is the observed \textbf{misalignment} between human rationales generated by the model (Alignment~\ref{user_study_alignment}), which directly contextualizes the ``syntax bias’’ identified in $RQ_{1.2}$.

This misalignment indicates that the model’s reliance on shallow syntactic heuristics (Sec.\ref{result:distribution}) is not simply a technical curiosity but a \textbf{meaningful divergence from expert reasoning}. Such divergences are precisely where trust becomes vulnerable. Previous work shows that misplaced trust can lead to failures such as data leakage or security risks\cite{thorneInterplay}, while a complete absence of trust can inhibit innovation~\cite{balayn2024empiricalexplorationtrustdynamics}. Alignment therefore sits at the center of fostering appropriate trust in \lmc~\cite{sun2024trustllmtrustworthinesslargelanguage, khati2025mappingtrustterrainllms}.

\codeRational’s primary contribution, in this context, is not to simply increase trust, but to help practitioners \textbf{calibrate} it. By revealing when the model’s reasoning is shallow (e.g., $RQ_{1.2}$) or diverges from human reasoning (Alignment~\ref{user_study_alignment}), the technique supports the development of \textbf{appropriate distrust}, which is conceptually distinct from a mere lack of trust~\cite{distrust2023}. Recent work shows that balanced trust and distrust enables more deliberate and critical engagement with \lmc, ultimately supporting improved decision-making and reducing error rates~\cite{distrust2023}.

\subsection{Threats to Validity}
\textit{Internal.} Some major threats might arise from our syntax taxonomy and sample selection in code completion and test case generation. To mitigate bias in sample selection, we utilized the open coding process in constructive grounded theory \cite{kathy2006}. Four authors reviewed each selected sample. We followed a similar approach to derive our taxonomy as well. We have also published all our data related to our taxonomy and sample selection in our online appendix ~\cite{anonymized_repo}.


textit{External:} Our results may be influenced by participant selection. We used purpose-sampling to recruit 86 experts and retained 37 complete responses (15 researchers and 22 developers or students). Although participants were experienced (for example, 92\% of Python users reported Intermediate or Expert proficiency), the group does not fully represent the broader developer population. All anonymized responses are released for transparency~\cite{anonymized_repo}.

A second threat concerns the absence of a baseline explanation method in our study design. Because the study was exploratory, our aim was to understand how practitioners evaluate \codeRational’s conceptual explanations rather than conduct a comparative assessment. A formal comparison is a natural direction for future work. Nevertheless, participant feedback suggests that the positive response was specific to our design; several noted that high-level \textbf{categories} were helpful and that token-level methods are \textbf{too fine-grained}, supporting the value of conceptual aggregation.

Our user study (\secref{sec:user_results}) examined perceptions of usefulness, and our future work (\secref{sec:conclusion}) outlines how these insights will translate into an interactive tool. However, we did not conduct a full task-based HCI evaluation demonstrating how \codeRational functions in a live development workflow. Such a study requires a tool implementation and a dedicated experimental design, which were outside the scope of this foundational work. As a result, claims regarding real-world in-workflow actionability should be interpreted with caution. This limitation is a primary direction for future work.

%% file: tabs/survey.tex
\begin{table}[!h]
\centering
\caption{User Study Results}

\label{tab:survery_metrics}

\resizebox{\linewidth}{!}{%
\setlength{\tabcolsep}{5pt} 
\renewcommand{\arraystretch}{1.3}

\begin{tabular}{cllllccc}
\multicolumn{1}{c}{} & \multicolumn{1}{c}{} &  & \multicolumn{1}{c}{} & \textbf{} & \multicolumn{3}{c}{\textbf{Results(\%answers)}} \\ \cline{6-8} 
\multicolumn{1}{c}{\multirow{-2}{*}{\textbf{\begin{tabular}[c]{@{}c@{}}Metric\\ ID\end{tabular}}}} & \multicolumn{1}{c}{\multirow{-2}{*}{\textbf{\begin{tabular}[c]{@{}c@{}}Downstream \\ Task\end{tabular}}}} &  & \multicolumn{1}{c}{\multirow{-2}{*}{\textbf{Use Case}}} & \textbf{} & \textbf{M. Agree} & \textbf{Neutral} & \textbf{M. Disagree} \\ \cline{1-2} \cline{4-4} \cline{6-8} 
 &  &  & Debugging a \lmc &  & 56.0 & 22.0 & 22.0 \\
 &  &  & Fine-tunning a \lmc &  & 89.0 & 11.0 & \cellcolor[HTML]{CBCEFB}{\ul 0.0} \\
 &  &  & Curating training/test/valid sets &  & 68.0 & 21.0 & 11.0 \\
 &  &  & Inferring causal relationship &  & 81.0 & 11.0 & 8.0 \\
\multirow{-5}{*}{\ref{user_study_usefulness}} & \multirow{-5}{*}{\textit{Model related}} &  & Interpreting \lmc output &  & 84.0 & 13.0 & 3.0 \\ \cline{1-2} \cline{4-4} \cline{6-8} 
 & \textit{Code Completion} &  & AST-based explanation informativeness &  & \cellcolor[HTML]{CBCEFB}{\ul 78.0} & 22.0 & \cellcolor[HTML]{CBCEFB}{\ul 0.0} \\
 & \textit{Test Case Generation} &  & Context-level explanation informativeness &  & \cellcolor[HTML]{CBCEFB}{\ul 79.0} & 11.0 & 10.0 \\
 & \textit{Code Completion} &  & AST-based representation readability &  & 52.0 & 27.0 & 21.0 \\
\multirow{-4}{*}{\ref{user_study_readability}} & \textit{Test Case Generation} &  & Context level representation readability &  & 59.0 & 14.0 & 27.0 \\ \bottomrule
\end{tabular}%

}
\end{table}

%% file: text/9_related_work.tex
\section{Related Work}\label{sec:related}


 In the context of Machine Learning and closer to the goals of our study, Lage \etal \cite{lage_human_2019} conducted controlled experiments on human subjects to study the degree to which the complexity of decision trees as surrogate models affects users' perceptions of response time, accuracy, and difficulty of use. Similarly, Poursabzi-Sangdeh \etal \cite{poursabzi-sangdeh_manipulating_2021} acknowledge the lack of consensus around defining, quantifying, or measuring the interpretability of ML models.


In ML, most research has focused on improving the plausibility and faithfulness of explanations such as LIME \cite{ribeiro_why_2016}, DeepLIFT \cite{shrikumar_learning_2019}, and Shapley values \cite{lundberg_unified_nodate}, to the best of our knowledge, our work is the first to qualitatively evaluate a feature importance-based interpretability method for code predictions. Incorporating human feedback into the evaluation of the desiderata and the faithfulness of the interpretability techniques is crucial \cite{chen_what_2022}. \codeRational has been evaluated through a user study to measure the utility of explanations for code generation from the user's perspective.

In the context of AI4SE, explanations for code generation have included counterfactual~\cite{cito_counterfactual_2022} and causal~ \cite{palacio_toward_2024}. In addition, methods based on \textit{self-attention} have been broadly adopted~\cite{mohankumar_towards_2020}. For example, Mohammadkhani \etal~\cite{mohammadkhani_explainable_nodate} propose an XAI method based on attention for three downstream tasks (\eg generation, refinement, and translation). However, attention-based explanations are unfaithful (\ie do not accurately reflect a model prediction) \cite{jacovi2020faithfully, jain_attention_2019, serrano_is_2019}. This issue arises because averaging attention weights across layers and heads is not a faithful score of individual token importance~\cite{brunner2020identifiability}. Basting \etal~\cite{bastings_elephant_2020} suggest that general input saliency methods are better suited to provide explanations than to attention. As demonstrated by Vafa \etal~\cite{vafa_rationales_2021}, \textit{rationales} contributes to more faithful and plausible explanations than gradient- and attention-based methods.

%% file: text/10_lessons_learned.tex
\section{Conclusion and Future Work}\label{sec:conclusion}

We introduced \codeRational, a framework for interpreting \lmc by moving beyond local token-level explanations to a global category-based analysis. Our work provides strong evidence for several key conclusions. First, \textbf{token-level explanations are insufficient for developer trust}. We found that the raw probabilities of token-level rationales are extremely weak (median < 0.012), but aggregating them into higher-level programming categories increases signal strength approximately six times (median $\approx$ 0.07), allowing for more coherent and human-aligned explanations. 

Second, the global view reveals that \textbf{LLMs exhibit complex, adaptive reasoning strategies}. Models dynamically shift from relying on syntactic heuristics and self-reliance (\eg \loops relying on \loops) to leveraging semantic cues from natural language as the prompt context changes. \codeRational exposes systemic reasoning patterns, such as \codeparrot’s strong reliance on shallow syntax over deeper semantics, which remain invisible to standard accuracy metrics.

These findings have important implications for future work. Our user study shows substantial misalignment between human and model reasoning (Jaccard similarity 0.074 for code completion, 0.306 for test generation), reinforcing the need for interpretability tools that clarify where model reasoning diverges. Participants found \codeRational’s explanations informative (78–84\% agreement), although graphical representations were harder to read (52–59\% agreement), underscoring that interpretability must be both faithful and usable. This presents an opportunity for HCI research to design clearer interfaces for global explanations.

We demonstrated \codeRational on two models to illustrate its ability to produce global code-based explanations. However, the rationalization process has $O(t \cdot p)$ complexity (or $O(p^2)$ in the worst case) and remains computationally costly, and open-model access is required for compatibility (\secref{sec:experiments}). These constraints highlight the broader challenge of building explanation frameworks that are both effective and scalable. While \codeRational is one instantiation, our broader aim is to promote global, developer-aligned interpretability as a guiding principle for future tools.

To move our findings from insights to action, we propose developing CodeQ as a VS Code extension. This tool can be designed for role-specific, on-demand support to avoid overwhelming users. For a \textbf{developer} (or "model user"), the tool provides simple, natural language insights triggered on-hover. For example, based on our finding in \secref{sec:user_results} that models over-rely on variable names, the tool could warn: \textit{CodeQ Insight: The model's generation was heavily influenced by the variable name \texttt{filter\_and\_return\_odds}. Consider adding type hints or a more explicit docstring to ensure semantic correctness.}'' For a \textbf{model debugger}, the tool would then provide the full heatmaps and statistical tables to diagnose this underlying syntactic bias.

\section{Acknowledgments}
This research has been supported in part by the NSF
CCF-2311469 and CCF-2346357. We also gratefully acknowledge support from Cisco Systems. 
The opinions, findings and conclusions
expressed in this work are solely those of the authors and do
not necessarily reflect the views of the sponsors.
